\documentclass[aps,reprint,amsmath,amssymb]{revtex4-2}

\usepackage{graphicx}
\usepackage{amsmath}
\usepackage{gensymb}
\usepackage{upgreek}
\usepackage{caption}
\usepackage{subcaption}
\usepackage{xcolor}
\usepackage[T1]{fontenc}

\begin{document}

\title{Using stochastic resonance of two-level systems to increase qubit decoherence times}

\author{Yujun Choi}
\affiliation{Department of Physics, Virginia Tech, Blacksburg, VA 24061, United States}
\affiliation{Virginia Tech Center for Quantum Information Science and Engineering, Blacksburg, VA 24061, United States}

\author{S.\ N.\ Coppersmith}
\affiliation{School of Physics, The University of New South Wales, Sydney, NSW 2052, Australia}

\author{Robert Joynt}\email{rjjoynt@wisc.edu}
\affiliation{Department of Physics, University of Wisconsin-Madison, Madison, WI 53706, United States}
\affiliation{Theoretical Sciences Visiting Program (TSVP), Okinawa Institute of Science and Technology Graduate University, Onna, 904-0495, Japan}

\date{\today} 

\begin{abstract}
Two-level systems (TLS) are the major source of dephasing of spin qubits in numerous quantum computing platforms.  In spite of much effort, it has been difficult to substantially mitigate the effects of this noise or, in many cases, to fully understand its physical origin. We propose a method to make progress on both of these issues.  When an oscillating field is applied to a TLS, stochastic resonance can occur and the noise spectrum is moved to higher frequencies. This shift in the TLS noise spectrum will increase the dephasing times of the qubits that they influence.  Furthermore, the details of this effect depend on the physical properties of the noise sources. Thus one can use qubit spectroscopy to investigate their physical properties, specifically the extent to which the TLS themselves possess quantum coherence. We find that it should be possible to determine the dephasing rate and the energy level separation of the TLS themselves in this way.      
\end{abstract}

\maketitle


\section{Introduction}

$1/f$ noise is a major source of decoherence in many quantum computing platforms \cite{paladino20141}.  It is generally believed to originate from two-level systems (TLS) whose fluctuations produce a random time-dependent field at the qubit \cite{muller2019towards, connors2019low, Freeman:2016p253108, Zimmerman:2014p405201, rudolph2019long, Takeda:2013p123113}.  To reduce the effect of $1/f$ noise, several experimental techniques have been proposed.  By far the best known is dynamical decoupling (DD) that compensates accumulated phase error by applying refocusing pulses to a qubit \cite{suter2016colloquium, Szankowski:2017p333001}.  DD can substantially increase the dephasing time of the qubit. However, it has the disadvantages of introducing more hardware and additional manipulation of the qubit.

Instead of driving the qubits to increase decoherence times, one may instead drive the noise sources that affect them. We shall call this noise source driving (NSD).  It is much less studied than dynamical decoupling.  One example of NSD is spin bath driving which has shown to be useful for nitrogen-vacancy center spin qubits in diamond.  This scheme applies a radio-frequency pulse or continuous wave to surface spins on the diamond, giving rise to partial decoupling of the dipolar interaction to the NV spin qubits \cite{de2012controlling, knowles2014observing}.  The spin bath driving shifts the noise power spectral density (PSD) from low to high frequencies, which increases the dephasing times \cite{bauch2018ultralong, bluvstein2019extending}. Experiment shows that this works well with both monochromatic and polychromatic driving fields~\cite{joos2022protecting}. This method is somewhat analogous to dynamical decoupling but it has the advantage that one does not manipulate the qubits - the carriers of quantum information - but rather the noise sources, which carry no useful information for quantum computing.  Hence we would expect that the technical requirements on the source of the driving fields are much less stringent for NSD than for DD.

Another NSD scheme for superconducting qubits has also been proposed. Theoretical study shows that the coherence times of a superconducting qubit can be increased by adding noise to the surrounding TLSs \cite{you2022stabilizing}.  It is noteworthy that the relaxation (decoherence) time of the qubit is closely related to the decoherence (relaxation) of the TLSs.  An experimental study exhibits a 23\% increase in the relaxation time of a transmon qubit on average when a DC electric field is applied to the TLSs \cite{lisenfeld2023enhancing}.

In the present work, we attempt to generalize and systematize the theory of NSD by using concepts from the established field of stochastic resonance (SR). 
SR as originally conceived is a phenomenon in which a charged particle moving back and forth randomly in a bistable potential is driven by an applied periodic oscillating electromagnetic field \cite{gammaitoni1998stochastic}.  It is certainly also possible to drive the system with sound waves, and this may be more practical in some circumstances \cite{Brehm:2017TLS,you2022stabilizing}. In the absence of driving, the dynamics are purely relaxational, typically leading to a  PSD that is Lorentzian.  When driven, one sees a modification of the PSD of the particle's motion that depends on the relation of the period of the oscillating field with the intrinsic relaxation rate.  SR differs from ordinary resonance in that it does not require that the system in question has a restoring force. Nevertheless, the modifications produced in the PSD can be dramatic, and this has important consequences for decoherence.

The theory of SR was developed originally in the weak-field limit \cite{mcnamara1989theory} and later in the strong-field limit \cite{shneidman1994weak, gammaitoni1995stochastic}.  A more general approach that is independent of external signal strength was also studied \cite{lofstedt1994stochastic}.   In recent years quantum stochastic resonance was observed experimentally in the tunneling process of electrons in a quantum dot \cite{wagner2019quantum} and in the spin dynamics of a single Fe atom under scanning tunneling microscope \cite{hanze2021quantum}.  The movement of weight to higher frequencies is maximized when the driving frequency is about half of the switching rate of the TLS \cite{gammaitoni1998stochastic}.

All of the above-cited theoretical work assumed essentially classical dynamics for the TLS.  The extension to quantum stochastic resonance where a spin-boson model was adopted instead of a classical particle in a bistable potential has also been examined \cite{lofstedt1994quantum, grifoni1996nonlinear}.  The traditional assumption is that relaxation times from the phonon bath are so short compared to the tunneling rate that it is justified to adopt a Markovian picture in which the transition rates do not reflect quantum coherence and therefore have no significant time dependence.  However, the actual evidence for this is slim, and of course, it may be true in some systems and not in others.  Hence it is important to have signatures of quantumness that can only be obtained using a theoretical framework in which coherent quantum effects can be present but which yet has a well-defined classical limit.    
The Lindblad equation satisfies these criteria, which is why we use it in this paper. 

This paper has two goals.  The first is to show that the shift in the PSD that SR produces can mitigate qubit decoherence, even in the presence of quantum coherence of the TLS.  The second is to show that NSD can reveal how much quantum coherence a TLS has, and of what type.

In Sec.~\ref{sec:Stochastic resonance} we review the theoretical background for SR.
In Sec.~\ref{sec:model}, the model for quantum TLS is introduced. We discuss the meaning of the parameters in the model and give some computational details. In Sec.~\ref{sec:singleTLS} we compute the shift in the noise PSD for a single TLS due to the driving and its consequences for the qubit dephasing time $T_{\phi}$.   In Sec.~\ref{sec:rabi} we deal with the case where the TLS dynamics are not purely relaxational.  In Sec.~\ref{sec:multiTLS} we extend this to multiple noise sources.
Finally, in Sec.~\ref{sec:discussion} we summarize the results for the mitigation of qubit decoherence and discuss the question of how much light NSD experiments can throw on the quantum nature of the noise sources.



\section{Stochastic resonance and Qubit Dephasing}
\label{sec:Stochastic resonance}
Classical SR is a phenomenon that occurs when a classical two-state system is driven by a sinusoidally oscillating external field. The dynamics of the undriven system are assumed to be purely relaxational so its PSD has a single peak at zero frequency.  When driven, the system also oscillates at higher frequencies and the height of the central peak decreases.

\subsection{Classical Model}
We establish notation and conceptual background in this section.  The treatment follows Refs.~\cite{mcnamara1989theory} and~\cite{lofstedt1994stochastic}. The two states of the system, (henceforth just the ``TLS'') are denoted by $s(t)=\pm 1$.  If at time $t$ we have that $s(t)=\pm 1$, then at time $t+dt$ the system has made a transition to $s(t)=\mp 1$ with probability $W_{\pm}(t) dt$.  We denote the probability that the TLS is in the state $\pm 1$ state at time $t$ by $P(\pm 1,t)$. Since $P(+1,t) + P(-1,t)=1$ it makes sense to define $s_z(t) = P(+1,t) - P(-1,t)$. The differential equation for $s_z(t)$ is

\begin{equation}
\label{eq:sz}
\frac{d s_z(t)}{dt} = - W(t) s_z(t) + \delta W(t). 
\end{equation}
Here $W = W_- + W_+$
and $\delta W = W_- - W_+$.
This equation can be solved exactly using the integrating factor method.  
Our main interest is in the correlation function $S(\tau)=\langle s(t+ \tau) s(t) \rangle$, where the angle brackets denote an average over initial conditions and over $t$.  The PSD is \begin{equation}
    S(\omega) = \frac{1}{2 \pi}
    \int_{ - \infty} ^{\infty} 
    d \tau
    e^{-i \omega \tau} S(\tau) .  
\end{equation}

We have that $S(-\omega) = S(\omega)$ so we will usually plot $S(\omega)$ only for $\omega > 0$.  The key point for the PSD is that since $S(\tau = 0) = \langle s_z^2(t) \rangle = 1$ we find that 
\begin{equation}
    \int_{ - \infty} ^{\infty} d \omega S(\omega) 
    = S(\tau = 0)=1.
\end{equation}
Thus the integral of the PSD is independent of the form of $W_{\pm}(t)$.  This implies that driving the TLS can move weight around in the integral of the PSD but the total weight (the value of the integral) is unaffected.

If there is no driving then $W_{\pm}(t) = W_{\pm}$ is independent of time and we find 
\begin{equation}
S(\omega)
=
\frac{2W/\pi}{W^2+\omega^2}
\end{equation}
in the simple case where $W_{+} = W_{-} = W/2$.  We have set $\delta W = 0$ here since it only affects the average value of $s(t)$ which is less interesting for our purposes.  

To drive the TLS we choose a driving frequency $\omega_d > 0$ and a driving strength $A$ and then set $W_{\pm} = W/2 + A\cos  \omega_d t $. One then finds \cite{mcnamara1989theory}
\begin{equation}
S(\omega) 
=
(1-R) \frac{2W/\pi}{W^2+\omega^2}
+\frac{R}{2}
[\delta(\omega - \omega_d) + 
\delta(\omega + \omega_d)].
\end{equation}
with 
\begin{equation}
R
=
\frac{A^2/2}{W^2+\omega_d^2}
\end{equation}

$R$ is the redistribution ratio that quantifies how much of the weight in the PSD has been moved from low frequencies of order $W$ to frequencies greater than $W$ (when $\omega_d >W$).  The sum rule for $S(\omega)$ implies that $0\leq R \leq 1$. 

When choosing $\omega_d$ there is a trade-off.  Large $\omega_d$ moves the delta-function at $\omega = \omega_d$ further to the right in a plot of $S(\omega)$ versus $\omega$, but the amplitudes of the delta functions decrease as $[\omega_d^2 + W^2]^{-1}$.

At higher order in $A$, harmonics appear in the PSD;
there are additional delta functions centered at $\omega_d, 2\omega_d, 3 \omega_d, ...$, and the noise spectrum is also modified \cite{lofstedt1994stochastic}.
In the Supplemental Material we show that when $\omega_d \gg W$, the functional form of the noise spectrum does not change significantly even when $A$ is large, so the $R$ ratio provides a good characterization of the shift in noise power induced by the driving.

\subsection{Effect on Qubit Dephasing}
\label{subsec:dephasing}

We now regard the TLS as a source of noise that acts on qubits. Our model of the TLS is very general, and capable of describing magnetic noise, for example.  For concreteness, however, let us suppose that the TLS is a fluctuating electric dipole $\mathbf{p}(t)$ so that the two states are $\pm \mathbf{p} = \mathbf{p}_0 s(t)$. Let the TLS be at the origin of coordinates and a charge qubit be at $\mathbf{r}$.  The operating frequency of the qubit is proportional to $E_i(\mathbf{r})$, the $i$th component of the electric field at the qubit.  The field of the TLS at $\mathbf{r}$ is $E_i(t) = -\partial V(t)/\partial x_i  $. with $V(t) = \mathbf{p}_0 \cdot \mathbf{r} \, s(t) / 4 \pi \epsilon_0 \epsilon_r r^2$, in a medium with relative permittivity $\epsilon_r$.  The dephasing time is determined by the correlation function
\begin{equation}
    S_E(\tau) = \langle E_i(t+\tau) E_i(t) \rangle
    \propto \langle s(t+\tau) s(t) \rangle
\end{equation}
which leads immediately to $S_E(\omega) \propto S(\omega)$.
For some further details about the application to a charge qubit, see the Supplemental Material.

We shall, for the most part, focus in this paper on qubits whose dominant decoherence channel is dephasing from low-frequency noise, and the associated decoherence time will simply be called $T_{\phi}$.  As long as $T_{\phi} \ll T_1$, we also have $T_{\phi} = T_2$. 

$S(\omega)$ controls dephasing as follows.  The off-diagonal elements of the qubit density matrix decay at short times according to $\exp(-t^2/T_{\phi}^2)$ with 
\begin{equation}
\label{eq:Tphia}
\frac{1}{T_{\phi}^2} 
=
C_0 \int_{0}^{\infty}
S(\omega) F(\omega T_{\phi}) d \omega.
\end{equation}
Here $F$ is a window function that selects for the frequencies in $S(\omega)$ that satisfy $\omega T_{\phi} \leq 1$.  For Gaussian noise we have $F(\omega T_{\phi}) = \sin^2(\omega T_{\phi}/2) / (\omega T_{\phi}/2)^2$. 
This is the form of $F$ that we use below to compute PSDs.  But for any kind of noise, $F$ has the same qualitative behavior.  An approximation that captures the essential physics is obtained by taking $F$ to be a step function, resulting in
\begin{equation}
\label{eq:Tphib}
\frac{1}{T_{\phi}^2} 
=
C_0 \int_{0}^{1/T_{\phi}}
S(\omega) d \omega.
\end{equation}
$T_{\phi}$ is limited only by the low-frequency part of the noise.  Of course, this is a very crude way to compute $T_{\phi}$ but in the absence of more details about the TLS there is little point in attempting to refine it.
The coupling constant $C_0$ depends on how the noise field is produced by the TLS (as just described for the example of an electric dipole), and how strongly the qubit couples to that field.  Some additional details for a charge qubit are again given in the Supplemental Material.

In Fig.~1 we give the graphical picture of how SR enhances $T_{\phi}$, using \ref{eq:Tphib}.  In the absence of driving the noise is in the central Lorentzian and the integral of that (together with $C_0$) gives $T_{\phi}$. When driven, the weight in the PSD moves above $1/T_{\phi}$, which is the upper cutoff in the integral, so $T_{\phi}$ is enhanced. 

\begin{figure}
     \centering
     \includegraphics[width=0.48\textwidth]{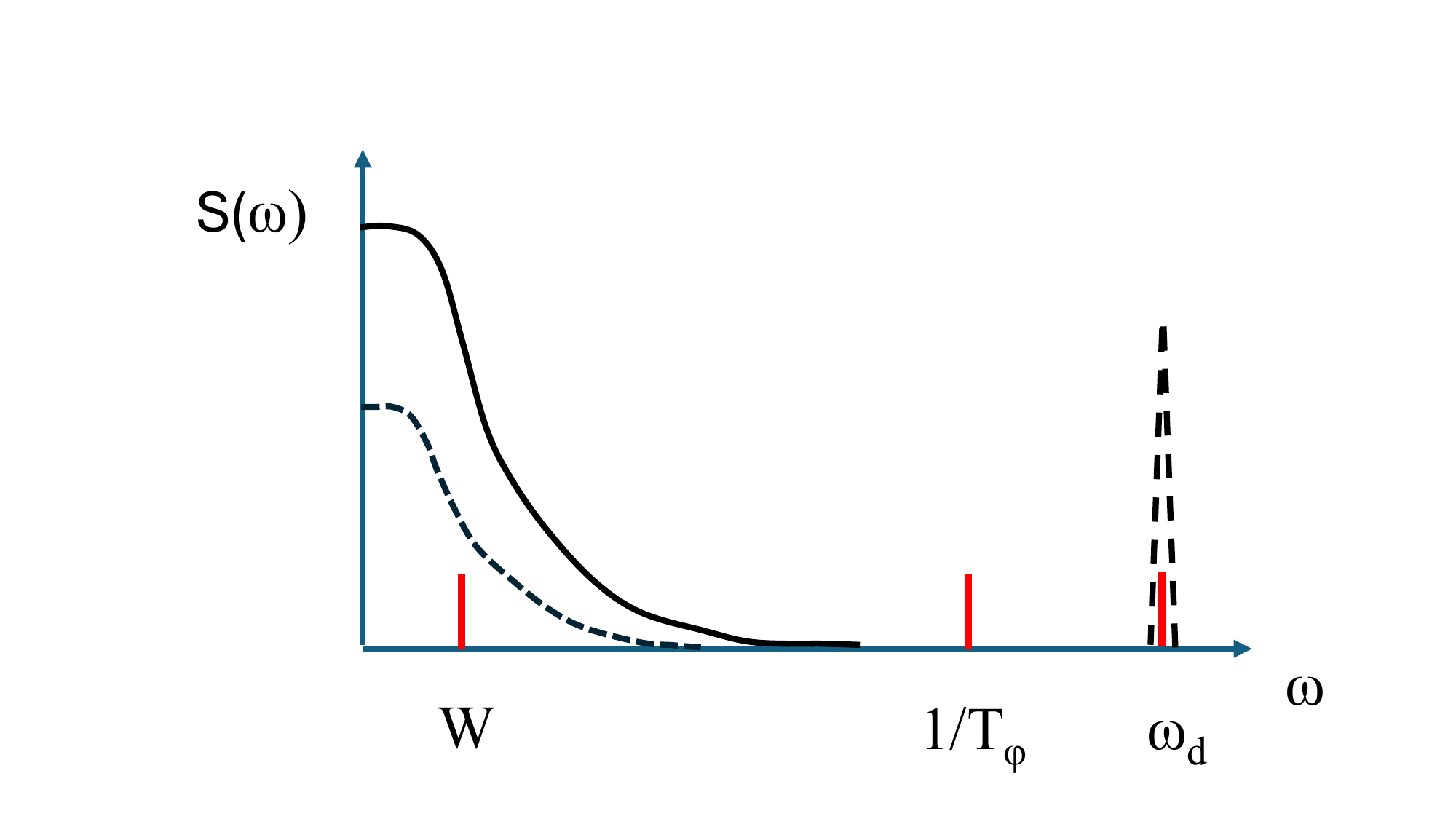}
     
     \caption[Strategy to use stochastic resonance to increase qubit dephasing times]  {Solid line is the PSD in the absence of driving, while the dashed line is the PSD of the driven TLS.  The peak at $\omega_d$ represents a $\delta$-function.  When $1/T_{\phi}>W$ and the TLS is driven at a frequency $\omega_d > 1/T_{\phi}$, then the weight in the integral in Eq.~\ref{eq:Tphia} to the left of the cutoff $1/T_{\phi}$ is reduced and $T_{\phi}$ increases.}
 \end{figure}

When the inequalities $W \ll 1/T_{\phi} \ll \omega_d$ hold so that the characteristic frequencies in the problem are well separated, then we can make definite statements about the enhancement of $T_{\phi}$.  Denoting the dephasing time in the absence (presence) of driving as $T_{\phi,0}$ ($T_{\phi,d}$),  Eq.~\ref{eq:Tphib} leads to 
\begin{equation}
    \frac{T_{\phi,d}}{T_{\phi,0}} = \frac{1}{\sqrt{1-R}}.
\end{equation}
This result holds also in the non-perturbative strong-driving regime, as long as $T_{\phi}$ does not exceed $1/W$. Once that happens, $R$ is no longer well-defined. In this regime $T_{\phi}$ can continue to increase, but that increase is no longer due to the movement of weight to frequencies greater than $W$.

\section{Quantum Model for noise sources}
\label{sec:model}
Since the origin of $1/f$ noise is not well known in most cases, we need to model the TLS as generally as possible.  Here we introduce a quantum model of a single TLS and compute its consequences for $T_{\phi}$.  


It is particularly important to include the possibility that the TLS is a quantum object in the sense that coherent superpositions of the two levels are allowed, unlike the model of Sec.~\ref{sec:Stochastic resonance}. The key question is when the classical SR behavior of the TLS is recovered when different parameters that govern the quantumness of the model are varied. 

\subsection{Lindblad Equation}
\label{subsec:lindblad}

A single TLS is an open quantum system coupled to phonons (and possibly other baths).  The dynamics are described by the Lindblad equation
\cite{manzano2020short}:
\begin{equation}
\frac{d\rho}{dt} = -i [H, \rho] + \sum_j \left( L_j \rho L_j^\dagger - \frac{1}{2}\left\{ L_j^\dagger L_j , \rho \right\} \right)
\end{equation}
where $\rho$ is the $2\times2$ density matrix for the TLS and $\hbar = 1$.

The off-diagonal elements $\rho_{01}$ and $\rho_{10}$ of the density matrix have no classical analog. 

The Hamiltonian is
\begin{equation}
\label{eq:hamiltonian}
H = \frac{1}{2} (\epsilon \sigma_z + \Delta \sigma_x) + (\alpha_z \sigma_z  + \alpha_x \sigma_x ) \cos \omega_d t
\end{equation}
where $\epsilon$ is the energy difference between the two levels, $\Delta$ is the tunneling amplitude, $\alpha_z$ and $\alpha_x$ are the driving strengths that modulate the energy difference and the tunneling, respectively.  In practice, $\alpha_z$ and $\alpha_x$ are couplings that depend on the properties of the qubit as well as the intensity of the driving field. 
 $\sigma_i$ are the Pauli operators and $\omega_d$ is the driving frequency.

The Lindblad operators are 
\begin{equation}
L_1 = \sqrt{\gamma}
\begin{pmatrix}
0 & 1 \\
0 & 0 
\end{pmatrix}, \,
L_2 = \sqrt{\kappa}
\begin{pmatrix}
0 & 0 \\
1 & 0 
\end{pmatrix}, \,
L_3 = \sqrt{\eta}
\begin{pmatrix}
1 & 0 \\
0 & -1 
\end{pmatrix}
\end{equation}
where $\gamma$, $\kappa$, and $\eta$ denote the relaxation, excitation, and pure dephasing rates, respectively. If $\epsilon < 0$ then $|0\rangle$ and $|1\rangle$ are the ground and excited states respectively.

We now change to variables whose physical meaning is more transparent.  As in Sec.~\ref{sec:Stochastic resonance}, we use $ s_z (t) = \rho_{00} (t) - \rho_{11} (t)$, and it is proportional to the electric dipole moment in the case of charged fluctuators that produce charge noise. The off-diagonal asymmetry is $p(t) = \rho_{01} (t) - \rho_{10} (t)$ and $q(t) = \rho_{01} (t) + \rho_{10} (t)$ is a phase measure. We also define the total relaxation rate: $\Gamma = (\gamma + \kappa)/2$, and the relative rate: $\lambda = \gamma - \kappa$.  

The detailed balance condition is $ 2\Gamma \tanh (\epsilon / 2 k_B T) =\lambda$, and we enforce this below in all our calculations, choosing $T = 0.1 $K.

Changing to these variables, we get the following coupled differential equations from the Lindblad equation:
\begin{equation}
\begin{aligned}
\frac{d s_z (t)}{dt} &= -2 \Gamma s_z (t) + \lambda - (\Delta + 2 \alpha_x \cos \omega_d t) p(t) \\
\frac{d p(t)}{dt} &=  (\Delta + 2 \alpha_x \cos \omega_d t) s_z (t)\\  &\quad - (\Gamma + 2\eta) p(t) - (\epsilon + 2 \alpha_z \cos \omega_d t) q(t) \\
\frac{d q(t)}{dt} &=  (\epsilon + 2 \alpha_z \cos \omega_d t) p(t) - (\Gamma + 2\eta) q(t).
\end{aligned}
\label{eq:coupled}
\end{equation}

\subsection{Classical Limit}
\label{subsec:classical}
Comparing Eqs.~\ref{eq:coupled} to Eq.~\ref{eq:sz} in the absence of driving we see that with the identifications $2 \Gamma = W$ and $\lambda = \delta W$ that the Lindblad equations we use reduce to the classical case when $p(t)=q(t)=0$ and $W_+$ and $W_-$ are independent of time.  Furthermore, even when $p(t)$ and $q(t)$ are included, if $\Gamma + 2 \eta$ is larger than the other parameters in the problem, both $p(t)$ and $q(t)$ decay quickly and again the Lindblad equation reduces to the classical limit after a short time (which is $T_2$).  

While large $\Gamma + 2 \eta$ leads to the classical limit, large $\epsilon $ and $\Delta$ increase the quantumness since they give rise to the Larmor precession about the axis in the $(x,y,z) = (\Delta,0,\epsilon) / \sqrt{\epsilon^2+\Delta^2} $.  This aspect of the quantum dynamics is absent in Eq.~\ref{eq:sz}, a classical equation.  The Larmor frequency $\sqrt{\epsilon^2+\Delta^2}$ is an additional frequency parameter in the problem, and thus we expect the PSD to have a complicated frequency structure, in contrast to the classical case, where only the harmonics of $\omega_d$ are involved.  We will investigate this in Sec.~\ref{sec:rabi}.

In addition, there are quantum effects in the structure of the Lindblad equations themselves in that the driving terms are not diagonal in the equation for $s_z$.  For driving to exist, $p(t)$ must be present.  This in turn implies the existence of off-diagonal elements in the density matrix, which means quantum coherence is present.

No closed-form solutions are known to exist for Eqs.~\ref{eq:coupled}, so we solve them numerically.  Ref. \cite{saiko2016dissipative} treats some analytic methods that describe the dynamics of a TLS under a strong driving field, but in a somewhat simpler model.

\subsection{Computational Details}
\label{subsec:computational}
The focus is on $s_z(t)$ because it is proportional to the dipole moment produced by the TLS.  Since $s_z (t) = \langle \sigma_z \rangle =$ Tr$[\rho(t)\sigma_z]$ is an expectation value, the correlation function for time duration $\tau$ can be regarded as $\pm s_z (\tau)$.  It does depend on the initial state,  $|0\rangle$ or $|1\rangle$, of an ensemble.  We are mainly going to deal with $\epsilon$ much smaller than $k_B T$, and we find that in the numerical calculations, the memory is sufficiently short that we can fix the initial state to $|0\rangle$. (For further details of the computations, see the Supplemental Material). Thus we can determine the quantity relevant for decoherence, which is $C_0 \int  s_z (\tau) e^{-i\omega \tau} d\tau$.  Therefore, we will look into the changes in $S(\omega)$ and $T_\phi$ caused by the driving field. To compute $S(\omega)$ we solve Eq. \ref{eq:coupled} numerically and then use the fast Fourier transform on $s_z(t)$. 

Since all of the interesting parameters may be expressed as frequencies when $\hbar = k_B = 1$, we set $\Gamma = 1$ so that the basic frequency is the switching rate of the TLS. 
 This fixes the choice of units for all of the quantities that appear in Eqs.~\ref{eq:coupled} as well as all other frequencies mentioned below.
 
 $s_z(t)$ is sampled with frequency $f_s = 10^4$ for a total simulation time $T_m = 10^2$ ($10^6$ points) for the single TLS calculations and $f_s = 10^3$ for $T_m = 10^3$ (again $10^6$ points) for multiple TLSs.  This is sufficient for our purposes since we will only look at the changes in $S(\omega)$ for $\omega < f_s/2$.

\section{Effect of the TLS dephasing time}
\label{sec:singleTLS}

The basic ideas of this work are already evident at the single TLS level.  We assume in this section that the energy levels of the TLS are degenerate.  This allows a simple comparison with classical SR since the existence of that phenomenon does not depend on the finiteness of the energy level separation.  In the quantum case, that separation leads to oscillatory behavior that will be investigated in Sec.~\ref{sec:rabi}.     

\subsection{Parameters}
\label{subsec:parameters}
We begin with the dephasing parameter $\eta = 1$, which means that dephasing is appreciable but not too strong. (Recall that the units are set by $\Gamma = 1$) 
 This choice is suggested by studies on high-frequency TLSs having similar or a few orders of magnitude smaller $\eta$ than $\Gamma$ \cite{lisenfeld2010measuring,lisenfeld2016decoherence}.
$k_B T/\hbar$ is normally much greater than the typical switching rates of TLS, and when we apply detailed balance we find $\lambda = 0$.
In this section, we also take $\Delta = 0$, and $\epsilon = 0$. 
 The two energy levels of the TLS are degenerate so there is no Larmor precession.  Precession effects will be treated in the next section.  Here the quantum effects arise only from the mixing of $s_z(t)$ with $p(t)$ and $q(t)$ as explained above.  We will vary the driving frequency $\omega_d$ and overall driving strength but enforce the relationship $\alpha_x = 0.5 \alpha_z$.  We may think of this relationship as off-axis driving since the precession axis of the driving field does not align with the $z$-axis that is associated with measurable quantities.

 \subsection{Power spectral density}
\label{subsec:psd1}



In general, we find that $\alpha_x$ is much more important than $\alpha_z$; there will be no driving effect if $\alpha_x \ll \alpha_z$.  For further discussion of their relative magnitudes, see the Supplemental Material. 


\begin{figure*}
\centering
\begin{subfigure}[t]{0.48\textwidth}
\includegraphics[width=\textwidth]{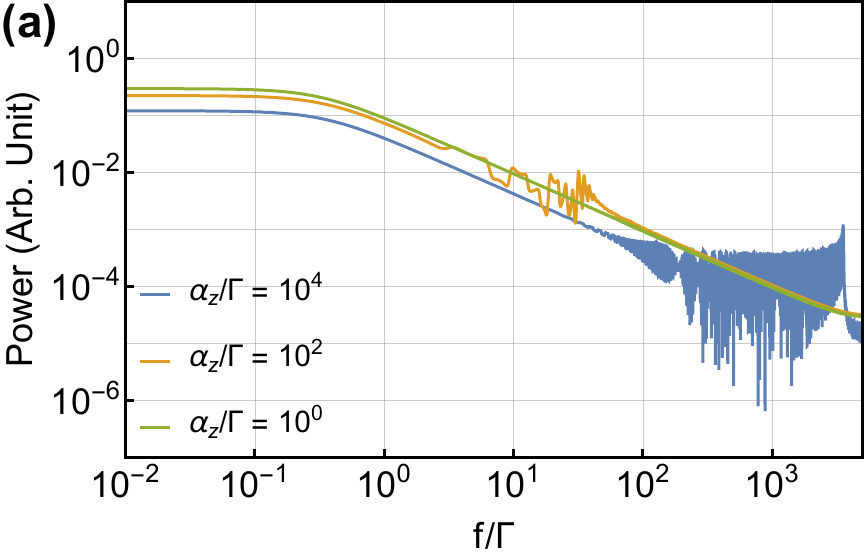}
\end{subfigure}
\hfill
\begin{subfigure}[t]{0.48\textwidth}
\includegraphics[width=\textwidth]{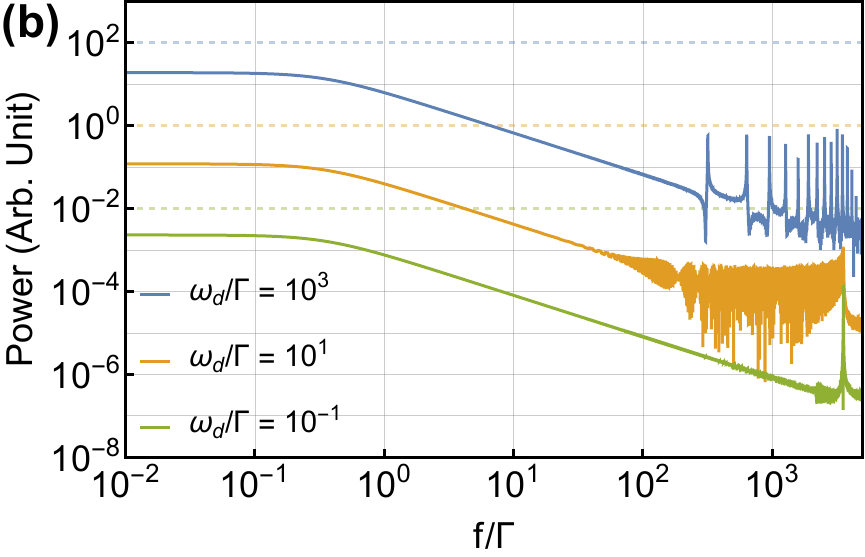}
\end{subfigure}

\caption[Shift of PSD for a single TLS]{\textbf{(a)} Shift of the PSD for a single TLS varying $\alpha_z$ with fixed $\omega_d = 10^1$. These PSDs correspond to the three boxed simulation points along the vertical axis in Fig.~\ref{fig:Tphi_single}. \textbf{(b)} Shift of PSD for a single TLS  varying $\omega_d$ with fixed $\alpha_z = 10^4$. These PSDs are offset by 2 orders of magnitude for clarity and correspond to the three boxed simulation points along the horizontal axis in Fig.~\ref{fig:Tphi_single}. Color-coded dashed lines mean the offset power at $\omega = 10^{-2}$ when the driving field is not applied.  All frequencies are in units of $\Gamma$, the width of the Lorentzian centered at $\omega = 0$.}
\label{fig:PSD_single}
\end{figure*}

 \begin{figure}
     \centering
     \includegraphics[width=0.48\textwidth]{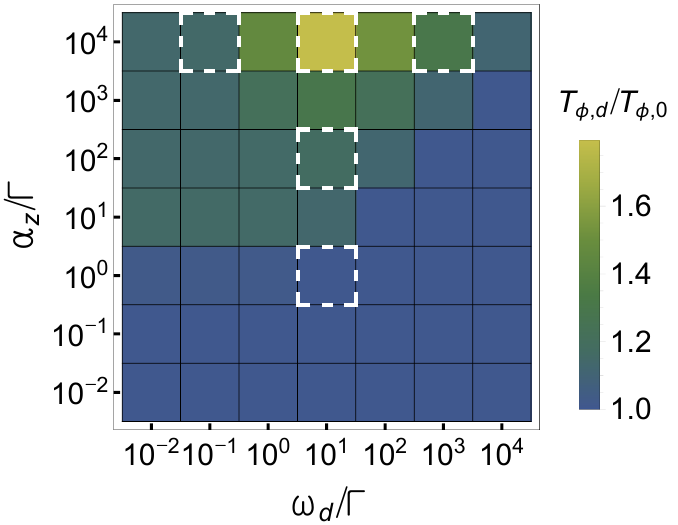}
     \caption[Increase in dephasing time of a qubit with a single TLS]{Increase in dephasing time of a qubit when a single TLS is driven by the oscillating electric field with strength $\alpha_z$ and frequency $\omega_d$. The dephasing time with driving field ($T_{\phi,d}$) is divided by that without driving field ($T_{\phi,0}$). Only 49 values of $T_{\phi,d}/T_{\phi,0}$ with the combination of 7 different strengths ($10^{-2}$ to $10^4$) and frequencies ($10^{-2}$ to $10^4$) are shown. White dashed boxes are for parameters used for the curves in Fig.~\ref{fig:PSD_single}.}
     \label{fig:Tphi_single}
 \end{figure}

Fig.~\ref{fig:PSD_single} gives the main results for the PSD.  

Fig.~\ref{fig:PSD_single}(a) shows the PSD when varying the driving strength $\alpha_z$.  The first important point is that the strength of the central Lorentzian does indeed decrease as $\alpha_z$ increases.  This means that this basic feature of classical SR is preserved in the quantum density matrix formalism.  The decrease is appreciable, quite visible on the log-log plot in the frequency range $\omega < 1$.  However, there are some differences between the results and what might be expected from classical SR.  We do not see a sharp peak at $\omega_d = 10$.  Instead, the weight for moderate driving is spread across a range of frequencies near $\omega_d$.  For strong driving, the weight moves up still further in frequency. 

Fig.~\ref{fig:PSD_single}(b) shows the PSD when the driving strength is strong ($\alpha_z = 10^4$) and fixed while $\omega_d$ is varied. The curves are offset, so the reduction in weight of the Lorentzian at low frequency is not easy to see.  The weight at high frequencies moves to the right as $\omega_d$ is increased.  This is in qualitative agreement with classical SR but again this weight is spread across a range of frequencies: there is no evidence of isolated sharp peaks. The structure at the far right in the small $\omega_d$ case is due to sampling error. Overall, the redistribution ratio $R$ is a function of both the driving strength and the driving frequency.

In practice, the decrease of $R$ may be mainly limited by practical considerations.  The frequency dependence is more interesting and shows the same trade-off mentioned in Sec.~\ref{sec:Stochastic resonance}.  At first, the shift of the weight ($R$ decreases) increases with frequency as the TLS attempts to follow the driving frequency, in somewhat the same fashion as a driven harmonic oscillator.  However, as the driving frequency continues to increase, the driving force averages out and the effect decreases.
We conclude that in the case where the Larmor precession is absent ($\epsilon = \Delta=0$) and intrinsic dephasing is moderate ($\eta=1$), quantum effects do not destroy the phenomenon of SR.  The main effect is to greatly complicate the frequency structure of the PSD at high frequencies. The single delta-function peak at $\omega_d$ converts to comb-like structures at higher frequencies.

\subsection{Qubit dephasing time}
\label{subsec:dephasing}
Any change in $S(\omega)$ will be reflected in the dephasing time $T_{\phi}$ of nearby qubits.  We use Eq.~\ref{eq:Tphia} with a Gaussian window function to compute $T_{\phi}$ as a function of driving strength and frequency.  To see the effects of driving, we compute the ratio of $T_{\phi,d}$, the dephasing time with driving, to $T_{\phi,0}$, that in the absence of driving.
The overall coupling strength $C_0$ is not so important here since we focus on the relative suppression of decoherence. However, to give an idea of experimentally relevant parameters in dimensionful units, if we take a coupling strength $C_0 = 10^{12}$ Hz$^2$ and a width of $W_0 = 1$ Hz, we will find $T_{\phi,0} = 1.00$ $\upmu$s, a typical value in spin qubit experiments. We emphasize once more that these are rough estimates since Eq.~\ref{eq:Tphia}
is an approximation.

In the simulations in Fig.~\ref{fig:Tphi_single}, $C_0 = 1 $ is used, resulting in $T_{\phi,0} = 2.31$.  The key point is that to have a significant increase in $T_\phi$, it is critical that $\omega_d > 1/T_{\phi,0}$.  If $\omega_d$ is smaller than this, the weight in the PSD may be shifted, but the shifted weight still contributes to the decoherence of the qubit, which is not what we want. Very large values of $\omega_d$ are also not favorable.  All of this is in agreement with classical SR.  

As shown in Fig.~\ref{fig:Tphi_single}, the increase in $T_\phi$ starts when $\alpha_z > \omega_d > \Gamma$, enhancing $T_\phi$ by a factor of up to 1.79.  There is no improvement in $T_\phi$ when $\alpha_z < \Gamma$ or $\alpha_z < \omega_d$.  This is in accord with the changes in $S(\omega)$ discussed above: the time scale of the driving strength should be shorter than the average relaxation time of the TLS to see the maximum effect.  

\subsection{Summary}
\label{subsec:summary}
\begin{figure}
     \centering     \includegraphics[width=0.48\textwidth]{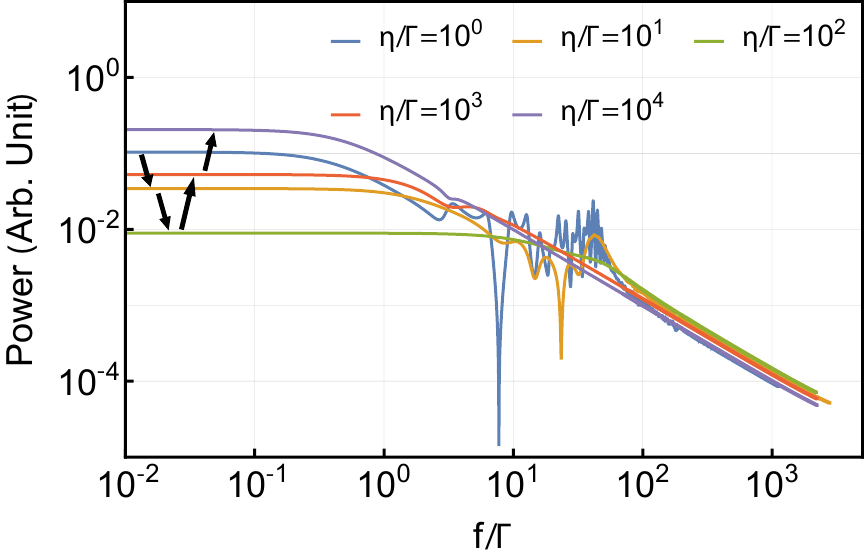}
     \caption{Power spectral density for different values of $\eta$, the dephasing parameter of the TLS.  This is the case that the energy levels of the TLS are degenerate: $\epsilon =0$. 
 The driving strengths are $\alpha_z = \alpha_x = 100$, and the driving frequency is $\omega_d = 10$.  The black arrows go in the direction of increasing $\eta$ and demonstrate the non-monotonic dependence of the redistribution ratio $R$ on the dephasing rate of the TLS.}
 \label{fig:PSD_eta}
 \end{figure}

Fig.~\ref{fig:PSD_eta} shows the results of varying the parameter $\eta$, which controls the dephasing, that is the decay of the off-diagonal parts of the density matrix $p$ and $q$.  Note that every curve crosses every other one: this is a consequence of the preservation of the total weight. 

$\eta$ can be thought of as a purely quantum parameter that complements $\Gamma$, which has a classical analog, as noted above.  In the $\hat{z}$ basis, both contribute to $T_2$ processes, while $\Gamma$ alone contributes to $T_1$ processes.

When $\eta$ is relatively small ($\eta/\omega_d \leq 1$) we see the effects of classical SR, as already noted.  Roughly speaking, the shift in weight can be followed by looking at the asymptotic value of the PSD at small $f$.  As $\eta$
increases, this value first decreases as the width of the Low-frequency Lorentzian increases, and the height decreases. The concomitant noisy behavior at higher frequencies is pronounced.  As $\eta$ increases still further, past $10 \, \omega_d$, decay of $p$ and $q$ is so rapid that no oscillations from the driving are possible. Exponential decay at a rate $\propto \Gamma$ results and there is no shift in weight.  The net effect is a non-monotonic shift in weight as a function of $\eta$, with the maximum around $\eta/\omega_d \approx 10$.  The black arrows in Fig.~\ref{fig:PSD_eta} indicate this.  This non-monotonicity is a purely quantum effect. We expect that it would manifest itself in an experiment as a non-monotonic behavior in $T_{\phi,d} / T_{\phi,0}$ as a function of the controllable parameter $\omega_d$.

\begin{figure*}
\centering
\begin{subfigure}[t]{0.48\textwidth}
\includegraphics[width=\textwidth]{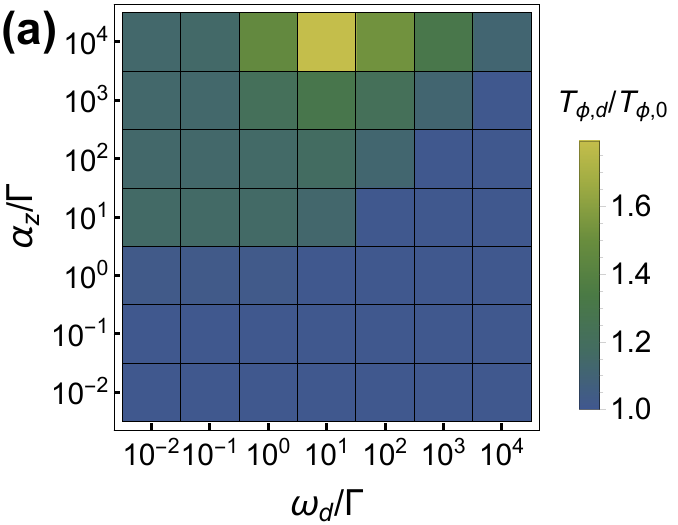}
\end{subfigure}
\hfill
\begin{subfigure}[t]{0.48\textwidth}
\includegraphics[width=\textwidth]{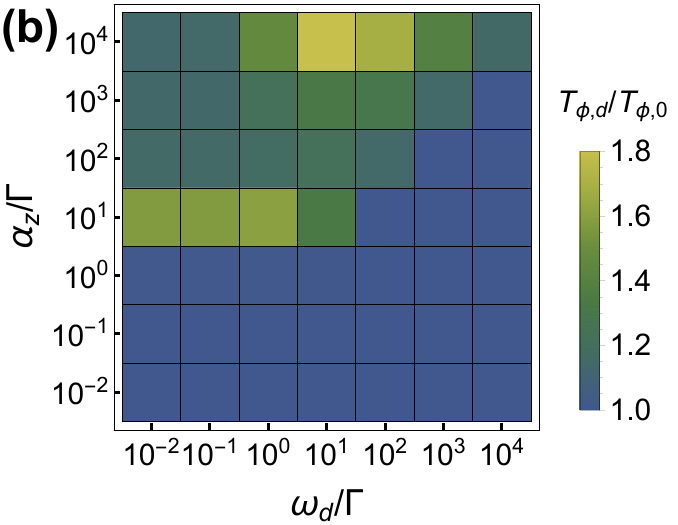}
\end{subfigure}

\caption{Enhancement of the dephasing time as a function of driving strength $\alpha_z$ and driving frequency $\omega_d$ of a qubit with energy level differences (a) $\epsilon = 0.1$ and (b) $\epsilon = 10$. 
 $\Gamma = 1$, $\eta = 1$, $\alpha_x = 0.5 \alpha_z$.  
 The slow precession in (a) does not change the overall pattern of the enhancement from the classical case. 
 The fast-precessing system in (b) shows a new peak in the enhancement when the precession frequency $\epsilon$ matches the scaled driving amplitude $\alpha_z/\Gamma$.}
\label{fig:Tphi_eps}
\end{figure*}

\section{Effect of Larmor precession of the TLS}
\label{sec:rabi}
We now introduce a nonzero energy level separation $\epsilon$ into the Hamiltonian. 
The two levels of a classical TLS also have different energies in general.  However, in the quantum case, this difference also implies a Larmor precession, an entirely nonclassical effect.  

We only consider the special case that $\Delta = 0$.  This means that the precession axis is the $z$-axis of the Bloch sphere.  Since we are looking at the correlation function of $s_z$, this means that the effect of the precession is somewhat indirect, as a rotation about the $z$-axis does not affect $s_z$.  This implies that in the absence of driving there is no peak in the PSD at $\Delta$. 

In Fig.~\ref{fig:Tphi_eps} we plot the $T_{\phi}$ enhancement as a function of the driving strength and the driving frequency for $\epsilon = 0.1$ (slow precession) and $\epsilon = 10$ (fast precession).   Slow precession is not enough to change the basic picture of essentially classical SR that we saw in the previous section. The enhancement still depends monotonically on the driving strength and there is the same peak in the enhancement as a function of driving frequency.  When the precession is fast the picture changes in one very important respect.  A peak in the enhancement as a function of driving strength appears at $\alpha_z \approx \epsilon $.

This peak can serve as an experimental signature of the energy level spacing made possible by a quantum effect.

\section{PSD and dephasing time of multiple TLSs}
\label{sec:multiTLS}

\begin{figure*}
\centering
\begin{subfigure}[t]{0.48\textwidth}
\includegraphics[width=\textwidth]{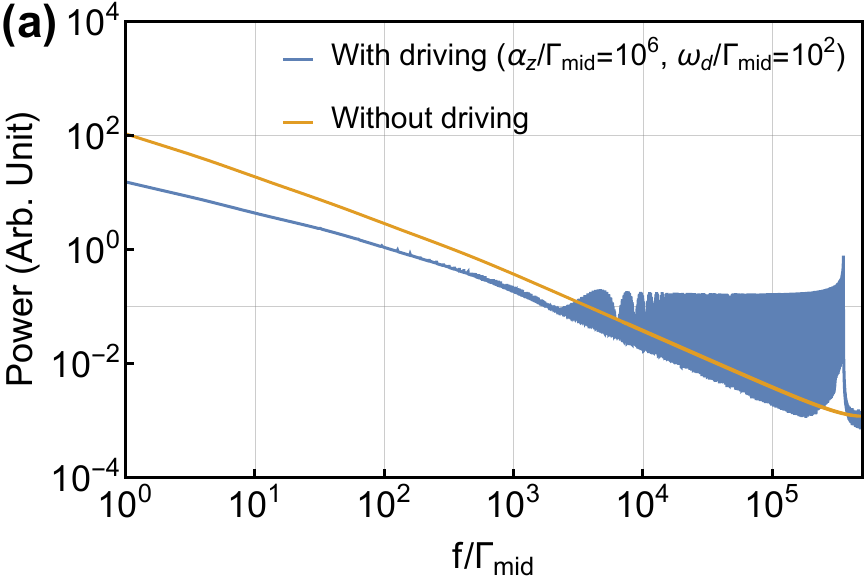}
\end{subfigure}
\hfill
\begin{subfigure}[t]{0.48\textwidth}
\includegraphics[width=\textwidth]{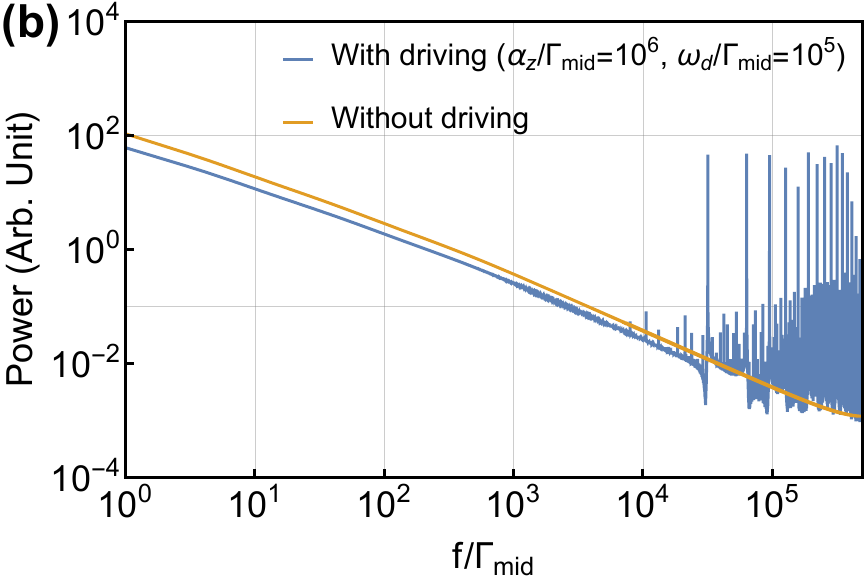}
\end{subfigure}

\caption[Shift of PSD for multiple TLSs]{Shift of the PSD of multiple TLSs as $\omega_d$ changes with fixed $\alpha_z = 10^3$. These PSDs correspond to the two simulation points in Fig.~\ref{fig:Tphi_multi}, respectively.  All frequencies are in units of $\Gamma_{mid}$, the switching rate in the logarithmic center of the seven TLS that produce the noise. (a) Here $\omega_d$ is in the range of the switching rates of the TLS. (b) Here $\omega_d$ is above the range of the switching rates of the TLS. Note the suppression of the low-frequency noise in panel (a).  This is nearly absent in panel (b).}
\label{fig:PSD_multi}
\end{figure*}

 \begin{figure}
     \centering
     \includegraphics[width=0.48\textwidth]{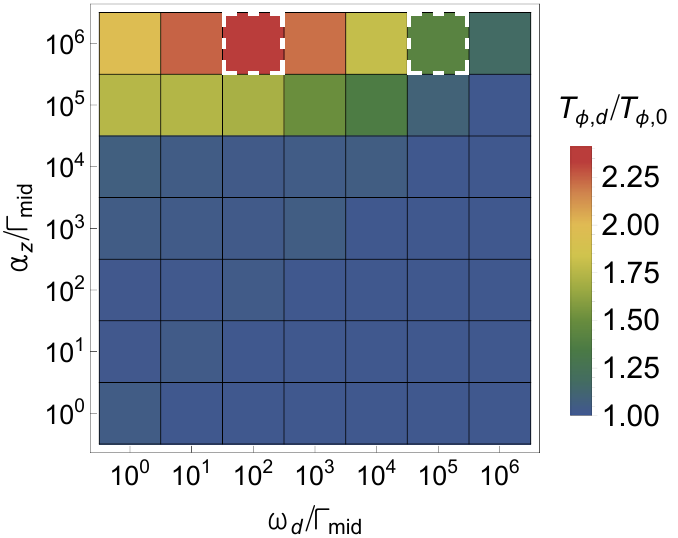}
     \caption[Increase in dephasing time of a qubit with multiple TLSs]{Increase in dephasing time of a qubit when multiple TLSs are driven by the oscillating electric field with strength $\alpha_z$ and frequency $\omega_d$. The dephasing time with driving field ($T_{\phi,d}$) is divided by that without driving field ($T_{\phi,0}$). 49 values of $T_{\phi,d}/T_{\phi,0}$ with the combination of 7 different strengths and frequencies are shown. $\Gamma_{mid}$ is the frequency in the logarithmic middle of the undriven PSD.  White dashed boxes are the simulation points used in Fig.~\ref{fig:PSD_multi}.}
     \label{fig:Tphi_multi}
 \end{figure}

The phenomenon of decoherence suppression has been established for the case of a single TLS.  In real systems many TLSs with a distribution of relaxation times are acting on the qubit, giving rise to PSDs that are generally not Lorentzian.  This suggests the question of whether a single driving source with a single driving frequency can still suppress qubit decoherence. 

We do not attempt a complete analysis of this question in this paper, which would involve optimizing the driving parameters for each type of PSD.  Instead, we merely try to settle the issue of whether NSD can mitigate the effects of the best-known PSD: $1/f$ noise.  This is produced by a set of TLS with a log-uniform distribution of Lorentzian widths.  Somewhat surprisingly, only one TLS per frequency decade is required to give a 1/f spectrum quite accurately. Motivated by this and computational convenience, we use 7 TLSs in this section. We set $\Gamma = \eta$ for each TLS. Their switching rates are set as $10^{-6}$, $10^{-5}$,$10^{-4}$,$10^{-3}$,$10^{-2}$,$10^{-1}$, and $10^{0}$.  Thus for this calculation, the unit of frequency is set equal to the fastest switching rate.  We also define $\Gamma_{mid}=10^{-3}$, which corresponds to logarithmic middle of the switching rates.  Other parameters are set to those used in Sec.~\ref{sec:singleTLS}.  Each TLS is simulated separately using the same method as above, and the individual power spectra $S_i(\omega)$ are added to give the total PSD, $S(\omega) = \sum_i S_i(\omega)$, which then yields the fluctuating field at the qubit.

The results are given in Fig.~\ref{fig:PSD_multi}, where we plot the total PSDs with fixed driving strength and two different driving frequencies. We include comparison plots with no driving.  The latter gives $1/f$ noise; the slight curvature in the plots is due to the fact that we only include 7 TLS, so at high frequencies the PSD has a $1/f^2$ behavior.    The changes in the PSD are more subtle than in the single TLS case.  The spectral weight can be moved to higher frequencies, but only in a limited range of $\omega_d$. This is a consequence of the fact that for the driving to have an effect on a TLS there needs to be an approximate match between the switching rate of the TLS and the driving frequency $\omega_d$.  This cannot be achieved for all of the TLSs when the noise is broadband since we limit ourselves to a single driving frequency.

It is clearly more effective to set 
$\omega_d$ in the frequency range of the TLSs, as in Fig.\ref{fig:PSD_multi} (a),
where $\omega_d = 10^{-1}$, rather than above that range, as in Fig.\ref{fig:PSD_multi} (b), 
where $\omega_d = 10^{2}$. 
The difference is most clearly understood by looking at small $f/\Gamma_{mid}$, 
where the suppression of the low-frequency noise noise is markedly suppressed in Fig.~\ref{fig:PSD_multi} (a) where $\omega_d/\Gamma_{mid} = 10^{2}$,
but much less so in Fig.~\ref{fig:PSD_multi} (b),
where $\omega_d/\Gamma_{mid} = 10^{5}$.
This is entirely in agreement with the above-mentioned importance of the matching of $\omega_d $ with switching rates.

The results for the change in the dephasing time for 7 TLSs are given in Fig.~\ref{fig:Tphi_multi}.  We see a qualitatively similar effect to that in Fig.~\ref{fig:Tphi_single}, but find the maximum value of $T_{\phi,d}/T_{\phi,0}$ to be about 2.43 when $\alpha_z = 10^3$ and $\omega_d = 10^{-1}$. This is quite intriguing because $\omega_d$ is smaller than $1/T_{\phi,0} = 2.27$ and inside the noise bandwidth ($10^{-6} - 1$). This strongly suggests that an optimal driving frequency should exist for a given $1/f$-like noise spectrum.  It is also most likely possible to do better using non-monochromatic driving.  However, the detailed optimization of the driving field is beyond the scope of this paper.

\section{Discussion}
\label{sec:discussion}

Driving the TLS noise sources does indeed change the PSD by moving weight to high frequencies where its effect on $T_{\phi}$ is reduced and qubit coherence is thereby enhanced. This may be a promising way to improve coherence in a system where dephasing from TLS is an important issue.  The weight shift occurs already at the classical level - it is present also in the theory of classical SR.  It does survive in the quantum case.  This can happen when the TLS is effectively classical because of its own rapid dephasing.  This is in fact the usual picture of a TLS, and in this paper we showed how this picture is recovered in one limit of the quantum equations.  

An important conclusion of the analysis was that the dependence of $T_{\phi}$ on the driving frequency is non-monotonic, with a peak in the enhancement when $\omega_d$ is comparable to or somewhat less than the TLS dephasing rate $\eta$. This provides a way for experiments to determine the rate of TLS dephasing, a quantity that is otherwise difficult to determine.

Precession of the TLS Bloch vector is another quantum effect 
that can affect the noise produced by the TLS. It comes from $\epsilon$, the difference in energy of the two levels.
When the precession is slow compared to the relaxation rate $\Gamma$ or to the driving strength parameter $\alpha_z$ then it does not affect the enhancement of qubit coherence.  However, when $\epsilon$ exceeds $\Gamma$ then there is a resonance when $\epsilon \approx \alpha_z$.  Observing this resonance as $\alpha_z$ is adjusted would give a measure of $\epsilon$. To see these effects on the PSD, it would be necessary that only one TLS dominates the noise at the qubit, which is of course not the typical situation.  However, specially designed experiments have succeeded in doing this \cite{ye2024characterization}.      

One objection to NSD is that it may be difficult in practice to drive the TLS electromagnetically without also driving the qubit at some level, with possible deleterious effects. For example, NSD may heat the sample.   This can induce a shift of resonant frequency for a spin qubit \cite{undseth2023hotter, choi2024interacting}. 
Sonic driving is also a possibility, though also here one may anticipate unwanted qubit driving, since elastic vibrations would, for example, modulate tunneling matrix elements in two-qubit gates that depend on the exchange interaction. However, the driving frequencies $\omega_d$ that are of interest are those that tend to be in the range of the frequencies of the slow noise that produce qubit dephasing, not the much higher frequencies characteristic of qubit operating frequencies.  This extreme mismatch implies that it should be possible to minimize the direct effect of the driving field on the qubits. We note also that dressed qubit protocols may be a useful tool for treating those effects that still remain. \cite{seedhouse2021quantum, hansen2021pulse}.

The physics of the TLS, the qubits, and their couplings clearly involve many parameters, and this paper covers only a small portion of the whole space.  For example, there is a regime where the qubit-TLS coupling $C_0$, in our notation, is large enough that the qubit can affect the dynamics of the TLS. 
 Such effects have been observed experimentally \cite{brox2011effects}.  The relatively simple equation we used for $T_{\phi}$ may not be valid in this strong coupling regime \cite{bergli2009decoherence}.  We considered only one off-axis direction for the driving field, and there may be interesting physics that was missed because of this limitation.  Temperature was only included by enforcing detailed balance between the upward and downward transitions. This problem can certainly benefit from further investigation.
 \vspace{5mm}




\section{Acknowledgments}
We thank Leah Tom for providing COMSOL simulation results in the Supplemental Material.  This research was partly conducted while visiting the Okinawa Institute of Science and Technology (OIST) through the Theoretical Sciences Visiting Program (TSVP).
This research was sponsored by the Army Research Office (ARO) under Awards No.\ W911NF-17-1-0274 and No.\ W911NF-W911NF-23-1-0110. 
The views, conclusions, and recommendations contained in this document are those of the authors and are not necessarily endorsed nor should they be interpreted as representing the official policies, either expressed or implied, of the Army Research Office (ARO) or the U.S. Government. The U.S. Government is authorized to reproduce and distribute reprints for Government purposes notwithstanding any copyright notation herein.
YC acknowledges support from the Army Research Office through Grant No. W911NF-23-1-0115.


\vspace{20mm}
\bibliography{main}

\end{document}